\documentclass[final,5p,times,twocolumn]{elsarticle}

\usepackage{lineno}
\usepackage{setspace}

\usepackage{amsfonts,amssymb,amsmath,bm,empheq}
\usepackage[dvipsnames]{xcolor}
\usepackage{balance}

\usepackage[colorlinks,citecolor=blue]{hyperref}
\usepackage{orcidlink}

\begin{document}

\begin{frontmatter}

\title{Reciprocal theorem for calculating the flow rate of oscillatory channel flows}

\author[PUME]{Shrihari D. Pande\fnref{fn1}\,\orcidlink{0000-0001-6962-8400}}
\author[Technion]{Evgeniy Boyko\fnref{fn1}\,\orcidlink{0000-0002-9202-5154}}
\author[PUME]{Ivan C.\ Christov\corref{cor1}\,\orcidlink{0000-0001-8531-0531}}
\ead{christov@purdue.edu}
\ead[url]{https://tmnt-lab.org}

\fntext[fn1]{These authors contributed equally.}
\cortext[cor1]{Corresponding author.}

\affiliation[PUME]{%
            organization={School of Mechanical Engineering, Purdue University},
            city={West Lafayette},
            postcode={47907}, 
            state={Indiana},
            country={USA}}

\affiliation[Technion]{%
            organization={Faculty of Mechanical Engineering, Technion -- Israel Institute of Technology},
            city={Haifa},
            postcode={3200003}, 
            country={Israel}}

\begin{abstract}
We demonstrate the use of the Lorentz reciprocal theorem in obtaining corrections to the steady flow rate due to flow oscillations in rigid channels. Starting from the unsteady Stokes equations, we derive the suitable reciprocity relation, assuming all quantities can be expressed as time-harmonic phasors. The auxiliary problem is the steady Hagen--Poiseuille flow solution, from which the reciprocal theorem allows us to calculate the first-order correction in the Womersley number to the steady flow rate in a straight rigid channel. We also consider nonuniform channels, specifically with variable height in the flow-wise direction, in which case the flow rate correction provides the leading-order effect of the interplay between the oscillations of the fluid flow and the given shape of the channel.
\end{abstract}




\end{frontmatter}




\section{Introduction}

Low-Reynolds-number oscillatory flows are widely encountered in microfluidics applications such as the mixing of fluids \cite{cola2006pulsed}, the generation of emulsions, and particle or cell separation \cite{dincau2020pulsatile}. Traditionally, the channels in these applications \cite{vedel2010pulsatile,vishwanathan2020generation,dincau2020pulsatile,recktenwald2021optimizing} are considered rigid, and it is of interest to determine the hydraulic impedance \cite{morris2004oscillatory} (i.e., the complex hydrodynamic resistance~\cite{kirby2010micro}) due to the oscillatory flow. 

Accurate modeling and design of such microsystems requires a quantitative understanding of the relationship between key integrated quantities, such as flow rate and pressure drop, in oscillatory flow. One approach to obtaining such relations is to solve the ``complete'' problem, meaning to solve the governing unsteady hydrodynamic equations. Another useful approach for finding integrated quantities (such as force, torque, or flow rate) is to exploit the Lorentz reciprocal theorem \cite{lorentz1896general}, which relates two different flow problems to each other (see, e.g., \cite{happel1983,masoud2019reciprocal} and the references therein). For example, in this way, a known solution of a ``simpler'' problem (termed the \emph{auxiliary} problem) can be used to obtain the flow rate of a ``harder'' problem \emph{without} the need for calculating its velocity and pressure fields.

This approach is particularly well-suited for low-Reynolds-number fluid mechanics applications such as those described above. For example, the reciprocal theorem has been used to predict the forces on and velocities of particles  \cite{zhang1998oscillatory,leshansky2004force,collis2017autonomous,kargar2021lift,fouxon2018fundamental,fouxon2023excess,zhang2023unsteady} and disks \cite{zhang1998oscillatory,usman2025small} in oscillating flow fields. Although many applications of the reciprocal theorem are to three-dimensional flows, two-dimensional (2D) external oscillatory flows caused by the motion of immersed surfaces can also be handled \cite{lin1987hydrodynamic,elfring2015anote}. In the present context, 2D flows are an idealization of 3D flows that are unconfined (or of large extent) in the spanwise direction perpendicular to the flow. Then, the lubrication approximation for slender flow geometries \cite{day2000lubrication} suggests that the velocity varies transversely and the pressure varies axially, and the reciprocal theorem has been used to obtain closed-form analytical expressions for flow rate--pressure drop relations of steady non-Newtonian flows through rigid \cite{boyko2021RT,boyko2022pressure} and deformable \cite{boyko2022flow,pande2025pressure,boyko2025interplay} channels. 

However, despite all the progress made, to the best of our knowledge, no application of the reciprocal theorem to \emph{oscillatory} channel flows has been presented to date. This observation motivates our study, which aims to show the reciprocal theorem's versatility and simplicity for studying the coupling between flow oscillations and channel shape.

To this end, in Sec.~\ref{sec:unsteady_stokes}, we formulate the problem of low-Reynolds-number flow in two-dimensional channels, and construct a suitable reciprocal theorem relating an oscillatory flow to a steady flow. Next, in Sec.~\ref{sec:lubrication}, we introduce the lubrication approximation and show how the reciprocal theorem can be used to obtain the leading-order correction (in the Womersley number) to the flow rate due to flow oscillations. The calculation is then implemented for a straight 2D rigid channel in Sec.~\ref{sec:straight_channel} and for a nonuniform 2D rigid channel, specifically with variable
height in the flow-wise direction, in Sec.~\ref{sec:nu_channel}. Finally, Sec.~\ref{sec:conclusion} concludes our study.

\section{Problem formulation and reciprocal theorem}
\label{sec:unsteady_stokes}

\begin{figure*}
    \centering
    \includegraphics[width=0.8\textwidth]{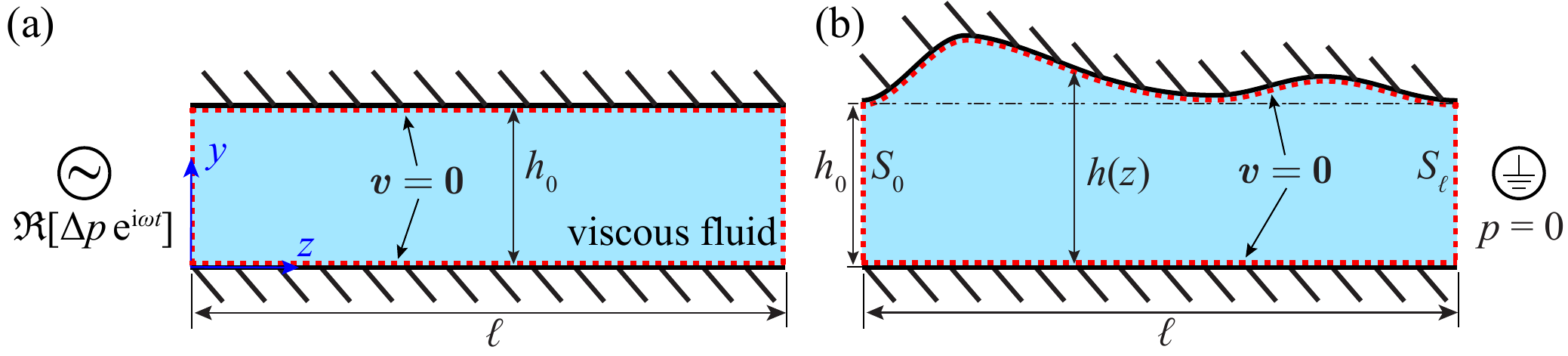}
    \caption{Schematic illustration of the two planar (2D) geometries considered: (a) a straight rigid channel and (b) a nonuniform rigid channel, specifically with variable height in the flow-wise direction. The straight channel has a height $h_{0}$, which is also the inlet/outlet height of the nonuniform channel. The integration domain $\mathcal{V}$ for the reciprocal theorem is the shaded interior of the channels. Its boundary is marked with bright dotted curves; $S_{0}$ and $S_{\ell}$ denote the inlet and outlet surfaces, respectively. The channels' axial length is $\ell \gg h$. An oscillatory viscous fluid flow is driven by an imposed pressure difference, $\Re[\Delta p \,\mathrm{e}^{\mathrm{i}\omega t}]$, with the outlet $(z=\ell)$ pressure setting the gage. The fluid velocity satisfies the no-slip condition along the stationary top and bottom walls of the channels (i.e., $\bm{v}=\bm{0}$ there). We are interested in determining the resulting flow rate $q$.}
    \label{fig:schematic}
\end{figure*}

Consider the incompressible oscillatory flow of a Newtonian fluid in the two 2D configurations shown in Fig.~\ref{fig:schematic}. The first configuration is a straight rigid channel of uniform height (Fig.~\ref{fig:schematic}(a)). The second configuration is a nonuniform rigid channel with variable height in the flow-wise direction (Fig.~\ref{fig:schematic}(b)). 
We assume that the axial length of the channels, $\ell$, is much greater than the channels' height $h$, i.e., we consider \emph{slender} geometries. The viscous fluid flow is driven by an imposed
oscillatory pressure difference with amplitude $\Delta p$ and frequency $\omega$, resulting in a velocity field $\boldsymbol{v}$ and a pressure distribution $p$. Our goal is to determine the flow rate $q$, and specifically how it depends on $\omega$ (or, dimensionless groups involving $\omega$), \emph{without} solving the oscillatory flow problem.

The corresponding governing equations are those of unsteady Stokes flow \cite{leal2007advanced,happel1983}:
\begin{equation}
    \boldsymbol{\nabla}\boldsymbol{\cdot}\boldsymbol{v}=0,\quad
    \boldsymbol{\nabla}\boldsymbol{\cdot}\boldsymbol{\sigma}=\rho\frac{\partial\boldsymbol{v}}{\partial t},\quad
    \boldsymbol{\sigma}=-p\boldsymbol{I}+2\mu\boldsymbol{E},
    \label{Continuity+Momentum RigChannel}
\end{equation}
where $\boldsymbol{I}$ is the identity tensor, $\boldsymbol{E}=\tfrac{1}{2}[\boldsymbol{\nabla}\boldsymbol{v}+(\boldsymbol{\nabla}\boldsymbol{v})^\top]$ is the rate-of-strain tensor, $\mu$ is the fluid's dynamic viscosity, and $\rho$ is the fluid's density. To retain the unsteady term but neglect the convective inertia, as in Eq.~\eqref{Continuity+Momentum RigChannel}, we introduce the Reynolds number $\mathrm{Re}={\rho v_c h_0}/{\mu}$ and the Womersley number $\mathrm{Wo} = h_{0}\sqrt{\rho\omega/\mu}$
\cite{womersley1955method,leal2007advanced} 
and require that $\mathrm{Re} \ll \mathrm{Wo}^2$ (or, equivalently, $v_c/\omega \ll h_0$, i.e., the flow oscillation amplitude is much smaller than the channel height) \cite{zhang1998oscillatory}. Here, $v_{c}$ is the characteristic axial velocity scale, and $h_0$ is either the rigid channel's height or the inlet/outlet height of the nonuniform channel.

We are only concerned with post-transient (``steady'') oscillations so that we can express the primary variables of the problem as \emph{phasors}:
\begin{subequations}
    \begin{align}
    \boldsymbol{v}(y,z,t) &= \Re[\boldsymbol{v}_{a}(y,z)\mathrm{e}^{\mathrm{i}\omega t}],\\
    p(y,z,t) &= \Re[p_{a}(y,z)\mathrm{e}^{\mathrm{i}\omega t}],\\
    \boldsymbol{\sigma}(y,z,t) &= \Re[\boldsymbol{\sigma}_{a}(y,z)\mathrm{e}^{\mathrm{i}\omega t}],
    \end{align}%
    \label{phasors2}%
\end{subequations}
where the quantities with an `$a$' subscript are complex-valued \emph{amplitudes}. Using Eqs.~\eqref{phasors2}, the linearity of the continuity and momentum equations~\eqref{Continuity+Momentum RigChannel}, and leaving the $\Re[\, \cdot \,]$ understood, we obtain 
\begin{equation}
    \boldsymbol{\nabla}\boldsymbol{\cdot}\boldsymbol{v}_{a}=0,\quad
    \boldsymbol{\nabla}\boldsymbol{\cdot}\boldsymbol{\sigma}_{a}=\rho\omega \mathrm{i}
    \boldsymbol{v}_{a},\quad
    \boldsymbol{\sigma}_{a}=-p_a\boldsymbol{I}+2\mu\boldsymbol{E}_{a}.
\label{Continuity+Momentum RigChannel phasor}
\end{equation}

Now, let $\hat{\boldsymbol{v}}$, $\hat{p}$, and $\hat{\boldsymbol{\sigma}}$ denote, respectively, the velocity, pressure, and stress fields corresponding to the \emph{steady} solution in the same channel, satisfying 
\begin{equation}
    \boldsymbol{\nabla}\boldsymbol{\cdot}\hat{\boldsymbol{v}}=0,\quad
    \boldsymbol{\nabla}\boldsymbol{\cdot}\hat{\boldsymbol{\sigma}}=\boldsymbol{0},\quad
    \hat{\boldsymbol{\sigma}}=-\hat{p}\boldsymbol{I}+2\mu\hat{\boldsymbol{E}}.
    \label{Continuity+Momentum Rigid RigChannel}
\end{equation}
Forming the product of the momentum balances in Eqs.~\eqref{Continuity+Momentum RigChannel phasor} and \eqref{Continuity+Momentum Rigid RigChannel},
with $\hat{\boldsymbol{v}}$ and $\boldsymbol{v}_a$, respectively, and using the usual vector identities, incompressibility, and the symmetry of the stress tensor \cite{masoud2019reciprocal,boyko2021RT}, we obtain 
\begin{subequations}
    \begin{align}
        \boldsymbol{\nabla}\boldsymbol{\cdot}(\boldsymbol{\sigma}_{a}\boldsymbol{\cdot}\hat{\boldsymbol{v}})-\boldsymbol{\sigma}_{a}\boldsymbol{:}\hat{\boldsymbol{E}} &= \rho \omega \mathrm{i}\boldsymbol{v}_a\boldsymbol{\cdot}\hat{\boldsymbol{v}},\\
        \boldsymbol{\nabla}\boldsymbol{\cdot}(\hat{\boldsymbol{\sigma}}\boldsymbol{\cdot}\boldsymbol{v}_a)-\hat{\boldsymbol{\sigma}}\boldsymbol{:}\boldsymbol{E}_a &= 0.
    \end{align}%
\end{subequations}
Subtracting these identities and using the stress tensors given in Eqs.~\eqref{Continuity+Momentum RigChannel phasor} and \eqref{Continuity+Momentum Rigid RigChannel}, yields
\begin{equation}
    \boldsymbol{\nabla}\boldsymbol{\cdot}(\boldsymbol{\sigma}_{a}\boldsymbol{\cdot}\hat{\boldsymbol{v}})-\boldsymbol{\nabla}\boldsymbol{\cdot}(\hat{\boldsymbol{\sigma}}\boldsymbol{\cdot}\boldsymbol{v}_a)=\rho \omega \mathrm{i}\boldsymbol{v}_a\boldsymbol{\cdot}\hat{\boldsymbol{v}}.
    \label{RT Os RigChannel 1}
\end{equation}

For \emph{either} configuration in Fig.~\ref{fig:schematic}, we can integrate Eq.~\eqref{RT Os RigChannel 1} over the entire fluid volume $\mathcal{V}$ bounded by the surfaces of the top and bottom walls, and the surfaces at the inlet and outlet ($S_{0}$ and $S_{\ell}$ at $z=0$ and $z=\ell$, respectively). Applying the divergence theorem leads to a reciprocal theorem in the form
\begin{multline}
    \int_{S_{0}}\boldsymbol{n}\boldsymbol{\cdot}\boldsymbol{\sigma}_{a}\boldsymbol{\cdot}\hat{\boldsymbol{v}} \,\mathrm{d} S 
    -\int_{S_{0}}\boldsymbol{n}\boldsymbol{\cdot}\hat{\boldsymbol{\sigma}}\boldsymbol{\cdot}\boldsymbol{v}_{a} \,\mathrm{d} S \\
    + \int_{S_{\ell}}\boldsymbol{n}\boldsymbol{\cdot}\boldsymbol{\sigma}_{a}\boldsymbol{\cdot}\hat{\boldsymbol{v}} \,\mathrm{d} S 
    - \int_{S_{\ell}}\boldsymbol{n}\boldsymbol{\cdot}\hat{\boldsymbol{\sigma}}\boldsymbol{\cdot}\boldsymbol{v}_{a} \,\mathrm{d} S 
    = \rho\omega \mathrm{i}\int_{\mathcal{V}}\boldsymbol{v}_{a}\boldsymbol{\cdot}\hat{\boldsymbol{v}} \,\mathrm{d} \mathcal{V}.
    \label{RT Os RigChannel 4}
\end{multline}
Here, $\boldsymbol{n}$ is the unit outward normal to $S_{0,\ell}$.
Note that the integrals over the bottom and top walls of the channel vanish because of the no-slip boundary conditions,  $\hat{\boldsymbol{v}}=\boldsymbol{v}_a=\boldsymbol{0}$, there. In passing, we note that Eq.~\eqref{RT Os RigChannel 4} is similar to Eq.~(3.13) of \citet{zhang2023unsteady}, who used the reciprocal theorem to obtain the force on an immersed sphere oscillating near a wall.

\section{Nondimensionalization and lubrication approximation}
\label{sec:lubrication}

Next, we introduce the dimensionless variables based on lubrication theory,
\begin{multline}
    T=\frac{t}{t_{c}},\quad Y=\frac{y}{h_{0}},\quad Z=\frac{z}{\ell},\quad V_{y}=\frac{v_{y}}{\epsilon v_{c}},\quad V_{z}=\frac{v_{z}}{v_{c}},\\
    Q=\frac{q}{q_c}, \quad P=\frac{p}{p_{c}}, \quad
    H=\frac{h}{h_{0}} ,
\label{Non-dimensional variables DefChannel}
\end{multline}
where $p_{c}=\mu v_{c}\ell/h_{0}^{2}$ is the characteristic lubrication pressure scale, 
$t_{c}=1/\omega$ is the characteristic time scale, and $\epsilon=h_0/\ell$ is the slenderness ratio of the channel, assumed to be small, i.e., $\epsilon\ll1$; this assumption is the \emph{lubrication approximation} \cite{leal2007advanced}.

We consider only planar geometries and analyze the problems per unit width. The flow is driven by an imposed pressure drop, which leads to $p_c=\Delta p \stackrel{\text{def}}{=} p|_{z=0} - p|_{z=\ell}$ so that the characteristic axial velocity scale is $v_c = h_0^2 \Delta p/(\mu \ell)$ and, consequently, the characteristic flow rate scale is $q_c = h_0 v_c = h_0^3 \Delta p/(\mu \ell)$. The regime of imposed flow rate is considered in~\ref{app:flow_rate_control}.

Using the dimensionless variables from Eq.~\eqref{Non-dimensional variables DefChannel} and performing a scaling analysis under the lubrication approximation (i.e., neglecting terms that are of $\mathrm{O}(\epsilon^2)$) as in \cite{boyko2021RT,boyko2022flow},  we obtain
\begin{subequations}
    \begin{align}
        \left.\boldsymbol{n}\boldsymbol{\cdot}\boldsymbol{\sigma}_{a}\boldsymbol{\cdot}\hat{\boldsymbol{v}}\right|_{z=0,\,\ell} &\simeq \mp\frac{\mu v_{c}^{2}\ell}{h_{0}^{2}}\left[-P_{a}\hat{V}_{z}\right]_{Z=0,\,1},\\ 
        \left.\boldsymbol{n}\boldsymbol{\cdot}\hat{\boldsymbol{\sigma}}\boldsymbol{\cdot}\boldsymbol{v}_a\right|_{z=0,\,\ell} &\simeq \mp\frac{\mu v_{c}^{2}\ell}{h_{0}^{2}}\left[-\hat{P}V_{z,a}\right]_{Z=0,\,1},
    \end{align}%
    \label{Scaling RigChannel}%
\end{subequations}
where the minus sign in Eqs.~\eqref{Scaling RigChannel} corresponds to $S_{0}$ and the plus sign corresponds to $S_{\ell}$.

Substituting Eqs.~\eqref{Scaling RigChannel} into Eq.~\eqref{RT Os RigChannel 4}, and using the outlet boundary condition $P_{a}|_{Z=1}=\hat{P}|_{Z=1}=0$, we obtain
\begin{multline}
    \int_{0}^{H(0)} \left[P_{a}\hat{V}_{z}\right]_{Z=0} \,\mathrm{d} Y-\int_{0}^{H(0)} \left[\hat{P}V_{z,a}\right]_{Z=0} \,\mathrm{d} Y \\
    = \mathrm{i} \mathrm{Wo}^2 \int_{0}^{1}\int_{0}^{H(Z)}V_{z,a}\hat{V}_{z} \,\mathrm{d} Y\mathrm{d} Z.
    \label{RT Os RigChannel 5}
\end{multline}
Noting that $P_{a} = P_{a}(Z)$ and $\hat{P} = \hat{P}(Z)$ under the lubrication approximation \cite{leal2007advanced}, and defining the flow rates,  $\int_{0}^{H(Z)}\hat{V}_{z} \,\mathrm{d}Y \stackrel{\text{def}}{=} \hat{Q}$ and $\int_{0}^{H(Z)}V_{z,a} \,\mathrm{d}Y \stackrel{\text{def}}{=} Q_a$, which are necessarily constant, 
Eq.~\eqref{RT Os RigChannel 5} yields the reciprocal theorem,
\begin{equation}
    \Delta P_{a} \hat{Q}
    = \Delta\hat{P}Q_a + \mathrm{i} \mathrm{Wo}^2 \int_{0}^{1}\int_{0}^{H(Z)}V_{z,a}\hat{V}_{z} \,\mathrm{d} Y\mathrm{d} Z,
\label{RT Os RigChannel 7}
\end{equation}
relating the oscillatory flow (`$a$' subscripts) of a Newtonian fluid in a 2D channel to its steady counterpart (`hats'). Note that Eq.~\eqref{RT Os RigChannel 7}, which is the key result of this work, is valid for \emph{both} oscillatory flow in a straight rigid channel and oscillatory flow in a nonuniform rigid channel with variable height in the flow-wise direction---the two configurations shown in Fig.~\ref{fig:schematic}.

Equation~\eqref{RT Os RigChannel 7} clearly shows that the dimensionless amplitude of the flow rate (or pressure drop) depends on $V_{z,a}$, and thus, generally, requires solving the complete oscillatory flow problem. However, in the weakly oscillatory limit, corresponding to $\mathrm{Wo}^{2}\ll1$, we can apply the reciprocal theorem \eqref{RT Os RigChannel 7} to obtain the \emph{leading-order oscillatory correction to the flow rate} using only the steady solution of the problem, as also observed in \cite{fouxon2018fundamental,zhang2023unsteady}. To this end, we expand the amplitude of the velocity into a perturbation series in the dimensionless parameter $\mathrm{Wo}^{2}\ll1$,
\begin{equation}
    V_{z,a} = \hat{V}_z + \mathrm{Wo}^{2}V_{z,a,1} + \mathrm{O}(\mathrm{Wo}^{4}),
    \label{Velocity RT Os RigChannel}
\end{equation} 
where $V_{z,a,1}$ is the first-order correction to the phasor amplitude of the axial velocity due to the oscillatory flow, and $V_{z,a,0}\equiv \hat{V}_z$ because the steady flow is the zero-Womersley-number limit of the oscillatory one (recall Eqs.~\eqref{Continuity+Momentum RigChannel} and \eqref{Continuity+Momentum RigChannel phasor}). The remaining calculations are thus all accurate to $\mathrm{O}(\mathrm{Wo}^{2})$. 


\section{Oscillatory flow in a straight rigid channel}
\label{sec:straight_channel}

Consider the straight rigid channel, shown in Fig.~\ref{fig:schematic}(a), for which $H=1$. The dimensionless solution of the \emph{steady} flow in this configuration, in a pressure-drop-controlled regime, is
\begin{equation}
    \hat{Q} = \frac{\Delta \hat{P}}{12},\quad
    \hat{V}_{z}(Y) = \frac{1}{2}\Delta \hat{P}(1-Y)Y.
    \label{Poiseuille RigChannel P control}
\end{equation}
Employing the expansion from Eq.~\eqref{Velocity RT Os RigChannel} and substituting $\hat{V}_z$ from Eq.~\eqref{Poiseuille RigChannel P control} into Eq.~\eqref{RT Os RigChannel 7}, we have 
\begin{equation}
    \begin{aligned}
    \Delta P_a \hat{Q} &= \Delta\hat{P} Q_a + \mathrm{i} \mathrm{Wo}^2 \int_{0}^{1}\int_{0}^{1}\hat{V}_{z}^{2} \,\mathrm{d} Y\mathrm{d} Z \\
    &= \Delta\hat{P} Q_a + \mathrm{i}\frac{1}{120}(\Delta\hat{P})^2\mathrm{Wo}^2.
    \end{aligned}
    \label{RT Os RigChannel 9}
\end{equation}
Finally, substituting $\hat{Q}$ from Eq.~\eqref{Poiseuille RigChannel P control} and imposing $\Delta P_{a}=\Delta \hat{P}=1$ (consistent with the pressure-drop-controlled regime), we obtain
\begin{equation}
    Q_a 
    = \frac{1}{12} \left( 1 - \mathrm{i} \frac{1}{10} \mathrm{Wo}^{2}  \right) .
    \label{RT Os RigChannel 10}
\end{equation}

\begin{figure}[t]
    \centering
\includegraphics[width=0.4\textwidth]{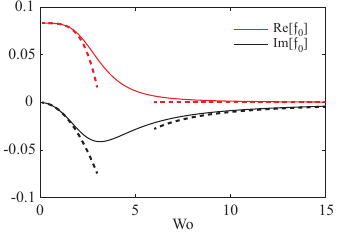}
    \caption{The real and imaginary parts of the function $\mathfrak{f}_0(\mathrm{Wo})$ from Eq.~\eqref{eq:f0_Wo}, which relates the (complex) amplitudes of the dimensionless flow rate and pressure drop for oscillatory flow in a straight 2D channel. Dark (black) and light (red) dashed curves represent the small- and large-$\mathrm{Wo}$ asymptotic expressions given in Eq.~\eqref{eq:f0_Wo}.}
    \label{fig:f0_and_f1}
\end{figure}

Oscillatory flow in a straight 2D channel has an exact solution, which allows us to ascertain the validity of our result in Eq.~\eqref{RT Os RigChannel 10}. Therefore, we review this exact solution. This flow is unidirectional such that $\boldsymbol{V} = V_z(Y,T) \boldsymbol{e}_z$ \cite{leal2007advanced}. Introducing the phasors from Eqs.~\eqref{phasors2} and the dimensionless variables from Eq.~\eqref{Non-dimensional variables DefChannel}, Eqs.~\eqref{Continuity+Momentum RigChannel phasor} reduce to 
\begin{equation}
    \frac{\partial^2 V_{z,a}}{\partial Y^2}-\frac{\partial P_a}{\partial Z} = \mathrm{Wo}^2\mathrm{i} V_{z,a}, \quad
    \frac{\partial P_a}{\partial Y}=0,
    \label{eq:lubrication_momentum_nondim}
\end{equation}
subject to no slip along the channel walls, $V_{z,a}(Y=0)=V_{z,a}(Y=1)=0$. The flow is driven by a time-harmonic, oscillatory pressure at the inlet with the gage pressure fixed at $Z=1$ (the outlet). Since the flow is unidirectional, and $V_{z,a}$ does not depend on $Z$, while $P_a$ does not depend on $Y$, it follows that \cite{leal2007advanced}:
\begin{equation}
    -\frac{\partial P_a}{\partial Z}=\Delta P_a \quad \Longleftrightarrow \quad P_a(Z)=\Delta P_a (1-Z).
    \label{eq:osc_pressure_nondim}
\end{equation}
Now, it is straightforward to show that the solution to Eqs.~\eqref{eq:lubrication_momentum_nondim}, taking into account Eq.~\eqref{eq:osc_pressure_nondim}, is
\begin{equation}
    V_{z,a}(Y) = \frac{1}{\mathrm{i}  \mathrm{Wo}^2}\left[1-\frac{\cos\left(\mathrm{i}^{3/2} (1-2Y) \mathrm{Wo}/2\right)}{\cos\left(\mathrm{i}^{3/2} \mathrm{Wo}/2 \right)}\right] {\Delta P_a}.
    \label{eq:vz_osc_channel}
\end{equation}
This form of the solution and its $\mathrm{Wo}$ expansion is also given by \citet[Sec.~11.7]{panton2013} (see also \cite[\S24]{LL87f} for a similar but not identical discussion). From Eq.~\eqref{eq:vz_osc_channel}, the dimensionless volumetric flow rate's amplitude is 
\begin{equation}
    Q_a = \int_0^1 V_{z,a}(Y) \,\mathrm{d}Y = \mathfrak{f}_0(\mathrm{Wo}) {\Delta P_a},
    \label{eq:flow_rate_chan}
\end{equation}
where, for convenience, we have defined the complex-valued function
\begin{equation}
\begin{aligned}
     \mathfrak{f}_0(\mathrm{Wo}) &\stackrel{\text{def}}{=} \frac{1}{\mathrm{i} \mathrm{Wo}^2} \left[ 1 - \frac{1}{\mathrm{i}^{3/2}{\mathrm{Wo}}/2} \tan\big(\mathrm{i}^{3/2}\mathrm{Wo}/2\big) \right] \\
     &\simeq
     \begin{cases}
        \displaystyle
        \frac{1}{12} - \mathrm{i}\frac{1}{120}\mathrm{Wo}^2 + \cdots, &\quad\mathrm{Wo} \ll 1,\\[3mm]
        \displaystyle
        0 - \mathrm{i}\frac{1}{\mathrm{Wo}^2} + \cdots, &\quad\mathrm{Wo}\gg1.
    \end{cases}
\end{aligned}
\label{eq:f0_Wo}
\end{equation}
Figure~\ref{fig:f0_and_f1} shows the real and imaginary parts of $\mathfrak{f}_0(\mathrm{Wo})$; see also the related discussion in \cite{morris2004oscillatory,vedel2010pulsatile}.

Clearly, the result in Eq.~\eqref{RT Os RigChannel 10} is in agreement with Eq.~\eqref{eq:flow_rate_chan} using the small-$\mathrm{Wo}$ expansion from Eq.~\eqref{eq:f0_Wo}.


\section{Oscillatory flow in a nonuniform rigid channel}
\label{sec:nu_channel}

Now, consider a nonuniform channel with a \emph{given} height variation $H(Z)$ in the flow-wise direction $Z$, as shown in Fig.~\ref{fig:schematic}(b). The dimensionless solution for the \emph{steady} flow in this configuration, in the pressure-drop-controlled regime, can be shown (by adapting the results from, e.g., \cite{tavakol2017extended}) to be
\begin{subequations}
    \begin{align}
    \hat{Q} &= \frac{\Delta \hat{P}}{12\int_{0}^{1}H(Z)^{-3} \,\mathrm{d} Z},
    \label{Poiseuille Q non-RigChannel P control}\\
    \hat{V}_z(Y,Z) &= \frac{\Delta \hat{P}}{2\int_{0}^{1}H(Z)^{-3}\,\mathrm{d} Z}\left[1-\frac{Y}{H(Z)}\right]\frac{Y}{H(Z)^2}.
    \label{Poiseuille V non-RigChannel P control}
    \end{align}%
    \label{Poiseuille non-RigChannel P control}%
\end{subequations}
Employing the expansion from Eq.~\eqref{Velocity RT Os RigChannel} and substituting $\hat{V}_z$ from Eq.~\eqref{Poiseuille V non-RigChannel P control} into Eq.~\eqref{RT Os RigChannel 7}, we have
\begin{equation}
\begin{aligned}
    \Delta P_a \hat{Q} &= \Delta\hat{P} Q_a + \mathrm{i}\mathrm{Wo}^2 \int_{0}^{1}\int_{0}^{H(Z)}\hat{V}_{z}^{2} \,\mathrm{d} Y\mathrm{d} Z \\
    &= \Delta\hat{P} Q_a + \mathrm{i}\frac{1}{120  \int_{0}^{1}H(Z)^{-3}\,\mathrm{d} Z}(\Delta\hat{P})^2\mathrm{Wo}^{2}\int_{0}^{1}\frac{1}{H(Z)}\,\mathrm{d} Z.
\end{aligned}
\label{RT Os non-RigChannel 9}
\end{equation}
Finally, substituting $\hat{Q}$ from Eq.~\eqref{Poiseuille Q non-RigChannel P control} and imposing $\Delta P_{a}=\Delta \hat{P}=1$, we obtain
\begin{equation}
    Q_a 
    =\frac{1}{12\int_{0}^{1}H(Z)^{-3}\,\mathrm{d} Z}\left[ 1 - \mathrm{i}\frac{1}{10} \mathrm{Wo}^{2} \int_{0}^{1}\frac{1}{H(Z)}\,\mathrm{d} Z \right] .
\label{RT Os non-RigChannel 11}
\end{equation}

Oscillatory flow in a nonuniform 2D channel under the lubrication approximation also has an analytical solution, though it does not appear in textbooks, which allows us to ascertain the validity of our result in Eq.~\eqref{RT Os non-RigChannel 11}. The lubrication approximation leads us again to Eq.~\eqref{eq:lubrication_momentum_nondim}, now subject to $V_{z,a}(Y=0)=V_{z,a}\big(Y=H(Z)\big)=0$; see, e.g., \cite{pande2023oscillatory} for a derivation. However, the pressure gradient is no longer constant in $Z$, thus Eq.~\eqref{eq:osc_pressure_nondim} becomes
\begin{equation}
    -\frac{\partial P_a}{\partial Z} = {\Delta P_a \mathcal{P}'(Z)}
    \quad \Longleftrightarrow \quad 
    P_a(Z) = \Delta P_a \mathcal{P}(Z) ,
    \label{eq:osc_pressure_nu}
\end{equation}
where primes denote differentiation with respect to the argument. The amplitude definition in Eq.~\eqref{eq:osc_pressure_nu} requires the normalization $\int_0^1 \mathcal{P}'(Z) \,\mathrm{d} Z = \mathcal{P}(1) - \mathcal{P}(0) = 1$. In principle, $\mathcal{P}(Z)$ is determined from the given shape variation $H(Z)$.

As in Sec.~\ref{sec:straight_channel}, it is straightforward to show that
\begin{equation}
    V_{z,a}(Y,Z) =  \frac{1}{\mathrm{i}  \mathrm{Wo}^2}\left[1-\frac{\cos\left(\mathrm{i}^{3/2} (1-2Y) \mathrm{Wo} \,H(Z)/2\right)}{\cos\left(\mathrm{i}^{3/2} \mathrm{Wo} \,{H}(Z)/2 \right)}\right] {\Delta P_a \mathcal{P}'(Z)}.
    \label{eq:vz_osc_channel_nu}
\end{equation}
From Eq.~\eqref{eq:vz_osc_channel_nu}, the volumetric flow rate's amplitude is found to be
\begin{multline}
    Q_a = \int_{0}^{H(Z)} V_{z,a}(Y,Z) \,\mathrm{d}Y\\
    = 
    \frac{H(Z)}{\mathrm{i} \mathrm{Wo}^2} \left[1 - \frac{1}{\mathrm{i}^{3/2} \mathrm{Wo}\,H(Z)/2}\tan\left(\mathrm{i}^{3/2} \mathrm{Wo}\,H(Z)/2\right)\right] {\Delta P_a \mathcal{P}'(Z)},
\label{eq:flow_rate_nu_chan}
\end{multline}
which must be constant since the cross-sectional area does not change in time. Then, for an imposed pressure drop, rearranging Eq.~\eqref{eq:flow_rate_nu_chan} and integrating from $Z=0$ to $Z=1$, we obtain
\begin{equation}
    Q_a \int_0^1 \frac{\mathrm{i} \mathrm{Wo}^2}{H(Z)}\frac{\mathrm{d}Z}{\left[1 - \frac{1}{\mathrm{i}^{3/2} \mathrm{Wo}\,H(Z)/2}\tan\left(\mathrm{i}^{3/2} \mathrm{Wo}\,H(Z)/2\right)\right]} = \Delta P_{a}.
\label{eq:nu_chan_dpa/qa}
\end{equation}

One can readily verify that the result in Eq.~\eqref{RT Os non-RigChannel 11} is in agreement with the small-$\mathrm{Wo}$ expansion of Eq.~\eqref{eq:nu_chan_dpa/qa}.

\section{Conclusion}
\label{sec:conclusion} 

In this communication, we revisited a basic problem in fluid mechanics using the Lorentz reciprocal theorem. Specifically, we derived a closed-form integral relation for calculating the flow rate in rigid 2D channels, with constant or axially varying heights, conveying oscillatory flows of incompressible Newtonian fluids. Our main result is Eq.~\eqref{RT Os RigChannel 7}, which relates the flow rates and pressure drops of steady and oscillatory flows (in 2D channels) to each other. For weakly oscillatory flow, we showed how the steady (Hagen--Poiseuille) flow solution can be used as the auxiliary problem in the reciprocal theorem to obtain the oscillatory flow rate up to $\mathrm{O}(\mathrm{Wo}^2)$. The flow rates obtained via the reciprocal theorem were verified by comparing to the low-Womersley-number asymptotic expressions stemming from the known analytical solutions (valid at any Womersley number). To the best of our knowledge, this admittedly simple but elegant calculation has not been done in the literature, highlighting the novelty of our approach.

Importantly, this approach allows one to bypass solving the momentum equation in obtaining the leading-order effect of flow oscillations. Thus, it would be straightforward to extend the present calculations for 2D channels to narrow axisymmetric tubes and three-dimensional (3D) channels of rectangular cross-section. Indeed, since the oscillatory pressure-driven flow field for the latter geometry is also known (see, e.g., \cite{morris2004oscillatory} and the references therein), one can find the leading-order effect of the flow oscillations and the cross-sectional aspect ratio of the 3D channel on the flow rate.

As an extension of this work, it would be of interest to adapt the reciprocal theorem to oscillatory flows in compliant conduits \cite{pande2023oscillatory,zhang2024elasto,huang2025oscillatory}.
Furthermore, as a future research direction, one could use the reciprocal theorem to explore how the rheological response of complex fluids, such as shear thinning and viscoelasticity, affects the oscillatory flow rate in both rigid and deformable channels.

\appendix



\section{Flow-rate-controlled regime}
\label{app:flow_rate_control}

For a flow-rate-controlled regime, we impose $q_c=q$; consequently,  $v_c=q/h_{0}$ and $p_c=\mu q\ell/h_{0}^{3}$.

\subsection{Straight rigid channel}

Now, the dimensionless solution of the \emph{steady} pressure-driven flow in a 2D straight channel is
\begin{subequations}
    \begin{align}
    \Delta \hat{P} &= 12\hat{Q},\\
    \hat{V}_{z}(Y) &= 6\hat{Q}(1-Y)Y,
    \end{align}%
    \label{Poiseuille RigChannel Q control}%
\end{subequations}
from which the calculation from Sec.~\ref{sec:straight_channel} can be redone. 
Specifically, imposing $Q_a = \hat{Q} = 1$ and substituting Eqs.~\eqref{Poiseuille RigChannel Q control} into Eq.~\eqref{RT Os RigChannel 7} and using Eq.~\eqref{Velocity RT Os RigChannel}, we obtain
\begin{equation}
    \Delta P_{a} 
    = 12 +\mathrm{i} \frac{6}{5} \mathrm{Wo}^{2} + \mathrm{O}(\mathrm{Wo}^4),
    \label{RT Os RigChannel 8}
\end{equation}
which is in agreement with the small-$\mathrm{Wo}$ expansion of ${1}/{\mathfrak{f}_0(\mathrm{Wo})}$ from Eq.~\eqref{eq:flow_rate_chan}.

\subsection{Nonuniform rigid channel}

Now, the dimensionless solution of the \emph{steady} pressure-driven flow in a 2D nonuniform rigid channel \cite{tavakol2017extended} is
\begin{subequations}
    \begin{align}
    \Delta \hat{P} &= \int_{0}^{1}\frac{12 \hat{Q}}{H(Z)^3} \,\mathrm{d} Z, \\
    \hat{V}_z(Y,Z) &= 6\hat{Q}\left[1-\frac{Y}{H(Z)}\right]\frac{Y}{H(Z)^2},
    \end{align}%
    \label{Poiseuille non-RigChannel Q control}%
\end{subequations}
from which the calculation from Sec.~\ref{sec:nu_channel} can be redone. 
Specifically, imposing $Q_a = \hat{Q} = 1$ and substituting Eqs.~\eqref{Poiseuille non-RigChannel Q control} into Eq.~\eqref{RT Os RigChannel 7} and using Eq.~\eqref{Velocity RT Os RigChannel}, we obtain
\begin{equation}
    \Delta P_a 
    = 12\int_{0}^{1}\frac{\mathrm{d} Z}{H(Z)^3} + \mathrm{i} \frac{6}{5} \mathrm{Wo}^2 \int_0^1 \frac{\mathrm{d} Z}{H(Z)}+ \mathrm{O}(\mathrm{Wo}^4),
\label{RT result NU-rigid channel}
\end{equation}
which is in agreement with small-$\mathrm{Wo}$ expansion of the left-hand side of Eq.~\eqref{eq:nu_chan_dpa/qa}.



\section*{CRediT authorship contribution statement}
\textbf{Shrihari D.\ Pande:} Investigation, Formal analysis, Validation, Writing - Original Draft, Visualization. \textbf{Evgeniy Boyko:} Conceptualization, Methodology, Investigation, Formal Analysis, Writing -- original draft, Writing -- review \& editing, Supervision, Funding acquisition. \textbf{Ivan C.\ Christov:} Conceptualization, Methodology, Investigation, Formal Analysis, Writing -- original draft, Writing -- review \& editing,  Supervision, Funding acquisition, Project administration.

\section*{Declaration of competing interest}
The authors declare that they have no known competing financial interests or personal relationships that could have appeared to influence the work reported in this paper.

\section*{Acknowledgements}
E.B.\ gratefully acknowledges support from the Israel Science Foundation (Grant No.\ 1942/23). S.D.P.\ and I.C.C.\ acknowledge partial support from the U.S.\ National Science Foundation (Grant No.\ CMMI-2029540) during the completion of this work.

\section*{Data availability}
No data was used for the research described in the article.

\setlength{\bibsep}{1ex}
\bibliographystyle{elsarticle-num-names} 
{\small
\bibliography{literature.bib}
}
\end{document}